\documentclass[10pt,a4paper]{article}
\usepackage{relsize}
\usepackage{amsmath}
\usepackage{amsfonts}
\usepackage{amssymb}
\usepackage{amsthm}
\usepackage{placeins}
\usepackage{changepage}
\usepackage{graphicx}
\usepackage{subfigure}
\usepackage{multirow}
\usepackage{chicago}
\usepackage{url}

\newcommand{\E}{\ensuremath{\mathbb{E}}}
\newcommand{\Mbb}{\ensuremath{\mathbb{Q}}}

\newcommand{\Qbb}{\ensuremath{\mathbb{Q}}}
\newcommand{\Qbbq}{\ensuremath{\Mbb}}

\newcommand{\Var}{\ensuremath{\text{Var}}}
\newcommand{\CVA}{\ensuremath{\text{CVA}}}
\newcommand{\FVA}{\ensuremath{\text{FVA}}}

\newcommand{\LGD}{\ensuremath{{L_{GD}}}}

\newcommand{\cov}{\ensuremath{\text{cov}}}
\newcommand{\var}{\ensuremath{\text{var}}}

\newcommand{\lB}{\ensuremath{\left(}}
\newcommand{\rB}{\ensuremath{\right)}}

\newcommand{\Ebb}{\ensuremath{\mathbb E}}

\newcommand{\sd}[1]{\ensuremath{\text{SD}(#1)}}

\theoremstyle{definition}

\begin{document}
\title{Model independent WWR for regulatory CVA \\and for accounting CVA and FVA}
\author{Chris Kenyon, Mourad Berrahoui and Benjamin Poncet\footnote{Contacts: chris.kenyon@mufgsecurities.com, mourad.berrahoui@lloydsbanking.com, benjamin.poncet@lloydsbanking.com. This paper is a personal view and does not represent the views of MUFG Securities EMEA plc (“MUSE”).   This paper is not advice.  Certain information contained in this presentation has been obtained or derived from third party sources and such information is believed to be correct and reliable but has not been independently verified.  Furthermore the information may not be current due to, among other things, changes in the financial markets or economic environment.  No obligation is accepted to update any such information contained in this presentation.  MUSE shall not be liable in any manner whatsoever for any consequences or loss (including but not limited to any direct, indirect or consequential loss, loss of profits and damages) arising from any reliance on or usage of this presentation and accepts no legal responsibility to any party who directly or indirectly receives this material.
This paper does not necessarily represent the view of Lloyds Banking Group.  No guarantees of any kind. Use at your own risk.}}
\date{08 October 2021 \vskip5mm Version 2.20}

\maketitle

\begin{abstract}
General wrong way risk (WWR) estimation is necessary for regulatory CVA capital and useful for pricing CVA and FVA.  We introduce a model independent method for calculating WWR and update the definition of WWR to deal with the lack of replication instruments (calibration data) transparently.   This model independent approach is extremely simple: we just re-write the CVA and FVA integral expressions in terms of their components and then calibrate these components.  This provides transparency between component calibration and CVA/FVA effect because there is no model interpretation in between.  Including funding in WWR means that there are now two WWR terms rather than the usual one.  Using a regulatory inspired calibration from MAR50 we investigate WWR effects for vanilla interest rate swaps and show that the WWR effects for FVA are significantly more material than for CVA.  This model independent approach can also be used to compare any WWR model by simply calibrating to it for a portfolio and counterparty, to demonstrate the effects of the model under investigation in terms of components of CVA/FVA calculations.
\end{abstract}

\section{Introduction}

General wrong way risk (WWR) estimation is necessary for regulatory CVA capital after 2023 (MAR50.32.7 and 50.41) and useful for pricing CVA and FVA.   WWR describes the situation where exposure increases and counterparty credit quality decreases.  Although more than thirty authors have proposed models for WWR (see Appendix 1) there is a data problem: credit volatility and credit-exposure correlation are not observable from liquid market instruments.  Credit default swap (CDS) options are only liquid for indices with a maximum expiry of less than one year\footnote{See any trading screen.}.  

Here we solve the WWR data issue by changing the definition of XVA WWR from WWR {\bf within} the XVA calculation, to WWR {\bf of} XVA where exposure, credit and funding are assumed independent.  This WWR {\bf of} XVA captures exactly the changes in existing independent CVA and FVA  calculations when WWR market events happen through the historical calibration choice which includes, for example, decrease of rates and increase in spreads in a credit crisis, or counterparty specific events, e.g. increase of the CDS spread coinciding with increase in the expected exposure.    We provide a model independent approach for calculation of both regulatory CVA WWR and accounting CVA and FVA WWR.   Our approach demonstrates that WWR depends on calibration choices, not modeling choices.

Calculating the WWR of XVA using historical estimates where market implied are not available aligns with market practice for correlation estimation, for example using the correlation between 15-year swaps of different currencies to drive the instantaneous correlation between driving processes for one or two factor models.  Our use is accurate because we will estimate, for example, the forward correlation between expected exposure and expected default probability using historical data and use the correlation at every tenor fosr which the correlation is estimated.  XVA depends on tenor, or terminal, volatilities and correlations we use the estimates directly, without any intermediate modeling.  This avoida any issues between instantaneous correlation and terminal (or tenor) correlation \cite{rebonato2005volatility}.   Furthermore we prove in Appendix 2 that the historically estimated correlations driving the WWR are identical for the two definitions, i.e. using exposure and default probability, or using their sampled expectations.

Model independence in our approach means that the calibration data directly define the parameters in the CVA and FVA expressions.  No additional parameterized models are required, although these parameterized models are the usual focus in WWR.  See \cite{massimi2019model} for a wider discussion of the model independent approaches.  This does not mean there are no assumptions, particularly in the calibration, and we list these in Table \ref{t:assump}. Putting the focus on WWR definition, calibration, and understanding calibration results, is necessary for WWR because there are no liquid instruments to calibrate to.

Our model independent approach is simply to re-write analytic expressions for CVA and FVA in terms of the two, or three, underlying elements and their correlations.  We continue re-writing in terms of simpler elements until they can be estimated directly from market, simulation, or historical data.  Once we have re-written the expressions the independent and dependent (WWR) parts of CVA and FVA are explicit, then we estimate the parameters within the re-written equations.  There is no model between these estimated parameters and CVA/FVA, the parameters directly affect regulatory and accounting results so the connection is transparent.  We propose estimation that is inspired by regulations (MAR50.10, CRE53.7, and d499) and use it to provide numerical results.

This paper has the following contributions: firstly we address the data issue in WWR and update the definition to align with available data and practice; secondly a model independent approach for estimating regulatory CVA for capital purposes: whilst Equation \ref{e:cva} is known, it has never been used for calculating WWR directly.  However, Equation \ref{e:cva} is insufficient for pricing because funding is absent.   Thus our third contribution is the extension of the model independent approach to accounting CVA and FVA including funding for both:  Equations \ref{e:acva2} and \ref{e:afva2} are new to XVA.  The contribution of using this model independent approach is that the connection between calibration data and effects on CVA and FVA for specific portfolios are directly visible, which we demonstrate allowing unprecedented insight: other techniques for WWR do not provide this level of portfolio transparency.  Finally we give a detailed assessment of the approaches' strengths and weaknesses for practical use.

\section{Model Independent WWR}

We develop WWR models by starting from an expression from CVA, or FVA, derived elsewhere \cite{green2016xva} or from MAR50 and simply rewrite it using basic arithmetic to reveal the WWR structure.  Once the WWR structure is sufficiently simple that we can estimate (aka calibrate) the parameters we can calculate the WWR impact on CVA and FVA.  

The model independent approach offers transparency between parameter calibration choices and their effects.   We also obtain structural information about the connection between counterparty portfolios and their WWR.

\subsection{WWR for regulatory CVA}

We first describe the regulatory CVA model, then re-write it in elementary terms and provide a list of the properties revealed.  Regulatory CVA is CVA calculated in accordance with MAR50, i.e. excluding any effect of own creditworthiness like funding effects. Parameters must be market-implied where possible and historically calibrated otherwise.

\subsubsection{Model}

A standard expression for CVA meeting MAR50 requirements is \cite{green2016xva}:
\begin{align}
\CVA =&  \int_0^T \E^\Qbb \left[\LGD(t) \lambda(t) D_{r,\lambda}(t) \max(0,\Pi(t))      \right]  dt  \label{e:cvaReg}\\
D_{r,\lambda}(t) =& \exp(-(r(t)+\lambda(t))) 
\end{align}
Notation is given in Table \ref{t:notation}.

\begin{table}
\begin{center}
\begin{tabular}{lp{8.5cm}}
\bf Symbol & \bf Meaning \\ \hline
$\Qbb$ & Risk-neutral measure \\
$r(t)$ & riskless short rate at time $t$ \\
$r_F(t) := r(t) + s_F(t)$ & funding short rate short rate at time $t$, equal to riskless plus funding spread \\
$\lambda(t)$ & hazard  rate for counterparty $C$ at time $t$ \\
\LGD & loss given default of counterparty $C$\\ 
$\Pi,\ \Pi^+$ & value of portfolio with counterparty, and exposure\\
Var($a$) & Variance of the distribution of the random variable $a$\\
SD($a$) & Standard Deviation of $a$\\
\hline
\end{tabular}
\caption{Notation}
\label{t:notation}
\end{center}
\end{table}

Given two random variables $a$ and $b$ it is elementary that
\begin{equation}
\Ebb[a b] = \Ebb[a] \Ebb[b] + \rho_{a,b} \sd{a} \sd{b}  \label{e:ab}
\end{equation}

Applying this to Equation \ref{e:cvaReg} with $a=\LGD(t) \lambda(t) D_{\lambda}(t) $ and $b=D_{r}(t) \max(0,\Pi(t))$ and using $\Pi^+ =\max(0,\Pi(t)) $ we get:
\begin{align}
\CVA 
        =&   \underbrace{\int_0^T  \E^\Qbbq_t[\LGD\lambda D_{\lambda}]\E^\Qbbq_t[D_{r} \Pi^+]dt}_\text{CVA, no WWR}  \nonumber \\
        &{}+   \underbrace{\int_0^T   \rho^\Qbbq_t \sqrt{ \Var^\Qbbq_t(\LGD\lambda D_{\lambda})  \Var^\Qbbq_t( D_{r} \Pi^+)} dt}_\text{CVA\ WWR only} 
        \label{e:cva}
\end{align}
For stochastic processes $\rho_t$  is the term structure of {\it terminal} correlation (not the instantaneous correlation) see \cite{rebonato2005volatility} Section 5.3.  Regulatory WWR for CVA is summarized in Table \ref{t:cva}.

An objection at this point might be that we have done nothing but re-write Equation \ref{e:cvaReg}.  This is exactly the simplicity of the model independent approach, we express WWR directly in terms that are applicable to CVA.  It is easy to understand the meaning of $\rho^\Qbbq_t$: correlation between forward default probabilities and forward exposures.

\subsubsection{CVA parameter estimation}
\label{s:estCva}

\begin{table}[t]
\begin{centering}
\begin{tabular}{llp{4.5cm}}
\bf term structure                         & \bf expression          & \bf source         \\  \hline
default-exposure correlation & $\rho^\Qbbq_t$                          & historical     \\
default variance             & $\Var^\Qbbq_t(\LGD\lambda D_{\lambda})$ & historical     \\
exposure variance            & $\Var^\Qbbq_t( D_{r} \Pi^+)$           & market implied, from CVA simulation    \\ \hline        
\end{tabular}
\caption{Parameter estimation for regulatory CVA WWR, for details see text.}
\label{t:cva}
\end{centering}
\end{table}

There are three items to estimate: exposure; default; and the terminal correlation between them, summarized in Table \ref{t:cva}.  Regulations state that market-implied estimation must be used where available, and if not available then historical estimation is permitted.
\begin{itemize}
\item Dynamics of underlyings for discounting and exposure: this is done as usual, with no change to standard (no-WWR) CVA setup.  Note that if there are any credit-dependent derivatives in the exposure then their dynamics are unchanged and exposure calculated as usual.
\item Dynamics of counterparty default probability and loss given default: historical estimation using their expectations observed on different days from the CDS spread curves.  As described in the introduction this is part of the updated definition of WWR to WWR of independent XVA rather than XVA with WWR.

 Hazard rates calibrated from CDS spreads are quite insensitive to discounting and this justifies the separation of hazard and (relatively fixed CDS)  exposure \cite{brigo2007interest} here.  As is usual in stochastic credit models we assume independence of default events and hazard rates.  For structured credit portfolios a more sophisticated approach may be of interest.

\item Term structure of default-exposure terminal correlation: this is estimated historically (see Numerical examples section for details) using a constant (current) portfolio against historical market data.  If there were standard traded CVA contracts, e.g. CCDS, for each counterparty then this could be market implied, but there are none.   Appendix 2 deals with correlation of sample expectations versus correlation, proving that they are the same.

We have two views on historical calibration: a regulatory-inspired view and an accounting view.  For WWR for regulatory CVA our view is inspired by looking at what is in regulations like MAR50.10, CRE53.7, and d499.  These regulations include stressed and a non-stressed historical periods so we do likewise: no stationarity assumption is required beyond what regulations assume.  For the accounting view of CVA and FVA we assume stationarity over the life of the portfolio.  Portfolio lifetimes with significant exposure can be twenty or more years so our view of stationarity is that there will be usual and stressed periods within that horizon.  The choice of which stressed periods to include in calibration is a matter for individual institutions, and complements stress testing. The difference between the two is that stress testing informs limits, whilst pricing informs PnL.

In terms of the correlation estimation method there is no dependence on the horizon.  The estimation method is the same for all horizons.  That is, if we can calculate exposures and default probabilities to calculate CVA using historical market data then we can use these elements to estimate terminal correlations. 

We take as our base case CVA calculated assuming independence between exposure and default, i.e. independent-CVA.  WWR CVA is relative to independent-CVA.  For WWR, from our parameter estimation, we are calculating how the independent-CVA changes as market and counterparty credit states change.  So for this use, estimating the correlation of sampled expected default probabilities versus sampled expected exposures is exact.

\end{itemize}

\begin{table}[t]
\begin{adjustwidth}{-2.0cm}{-2.0cm}
\begin{tabular}{lp{7cm}cp{5cm}}
\bf name & \bf assumption & \bf H & \bf comments \\ \hline
Reg-Style  & For regulatory CVA a regulatory-style calibration is appropriate  & No &  Applies to Regulatory CVA WWR.  Inspiration for Accounting CVA and FVA WWR        \\
Reg-Source & A regulatory-style calibration would include stressed and non-stressed periods as in MAR50  & No &   Stress periods occur roughly every 10 years so also relevant for Accounting WWR.        \\
Stress-stationarity & Future stress periods will look like past stress periods & No &  Current regulatory assumption in MAR50.10, CRE53.7, and d499.  Discussed for Accounting CVA and FVA in text. \\
Usual-stationarity & Future usual times look like current usual times  & Yes & Assumes no permanent changes to markets such as smile introduction  \\
1+2 Moments     & First and second moments exist exposure and default probability         & Yes  &  Practically this is always true and necessary. Some theoretical fat-tailed distributions have no moments     \\ \hline
\end{tabular}
\end{adjustwidth}
\caption{Assumptions within the CVA and FVA expression parameter estimation.  H column indicates whether this is typical for historical calibration.  Usual historical calibration to parameterized models have difficulty including stressed periods, but model independent approaches like historical VaR do not.  Stressed periods are also a feature of regulatory calculations.}
\label{t:assump}
\end{table}

The implicit and explicit assumptions in the parameter estimation are listed in Table \ref{t:assump}.

\subsubsection{Properties}

\begin{itemize}
\item Equation \ref{e:cva} is exact: this is trivially true as it is simply an elementary re-writing of CVA.  No assumptions are required for the CVA and FVA expressions on exposure, underlyings, default, or correlation beyond the existence of first and second moments.  The assumptions involved in estimation the parameters of these expressions are collected in Table \ref{t:assump}.
\item The terminal correlation term structure is deterministic.   $\rho^\Qbbq_t$ is {\bf not} stochastic. 
\item The terminal  correlation term structure is counterparty specific.
\item The re-writing cleanly separates the items that can be market implied (the expectations and exposure variances) from those that need historical estimation (terminal correlation and default variance). 
\item The re-write demonstrates that WWR (or RWR) is a basic, first order, property of CVA in the same way that $\Ebb[a b] = \Ebb[a] \Ebb[b] + \rho_{a,b} \text{SD}(a)\text{SD}(b)$.
\end{itemize}

\subsection{WWR for pricing and accounting CVA and FVA}

Pricing CVA and FVA usually include funding spreads and costs, thus there are three elements to consider: exposure; default; and funding, not two as in Regulatory CVA.  It is not surprising that there will now be two WWR terms rather than one.  We apply the same re-writing approach starting from CVA and FVA formulae analogous to the regulatory case:
\begin{align}
\CVA =&  \int_0^T \E^\Qbbq\left[ \LGD(t) \lambda(t) D_{r_F,\lambda}(t) \Pi^+(t) \right] dt  \label{e:acva1} \\
\FVA =&  \int_0^T \E^\Qbbq\left[ s_F(t) D_{r_F,\lambda}(t) \Pi(t) \right] dt  \label{e:afva1}
\end{align}
Given three random variables $\{a,b,c\}$ we can re-write the expectation of their product:
\begin{align}
\Ebb[a b c]  =& \Ebb[a] \Ebb[bc]  + \rho_{a,bc}  \sd{a} \sd{bc}  \\
                   =&  \Ebb[a] \Ebb[b] \Ebb[c] + \Ebb[a]   \rho_{b,c}  \sd{b} \sd{c} + \rho_{a,bc}  \sd{a} \sd{bc} \label{e:bc} \\
                   =&  \Ebb[a] \Ebb[b] \Ebb[c] + \Ebb[b]   \rho_{a,c}  \sd{a} \sd{c} + \rho_{b,ac}  \sd{b} \sd{ac}  \label{e:ac} \\
                   =&  \Ebb[a] \Ebb[b] \Ebb[c] + \Ebb[c]   \rho_{a,b}  \sd{a} \sd{b} + \rho_{c,ab}  \sd{a} \sd{ab}  \label{e:ab}
\end{align}
and taking $ \sd{ab}$ for example we can re-write it so as to separate functions of $a$ from functions of $b$:
\begin{align}
 \text{Var}(ab) =&  \rho_{a^2,b^2}\sd{a^2}\sd{b^2} + \Ebb[a^2]\Ebb[b^2] \nonumber \\
                   &{} - \left(  \rho_{a,b} \sd{a} \sd{b}  +  \Ebb[a] \Ebb[b]        \right)^2  \label{e:var1}
\end{align}
There are now three equivalent expressions for CVA and FVA depending on which re-write equation, \ref{e:bc}, \ref{e:ac}, or \ref{e:ab} we pick.  Note that by using Equation \ref{e:var1} we achieve a clean separation between market-implied items (expectations and exposure variances) and historically calibrated items (terminal correlations, default variances, and funding variances).   

\begin{align}
\CVA 
=&  \underbrace{\int_0^T  \E^\Qbbq_t[ \LGD \lambda D_{\lambda}]  \E^\Qbbq_t[ D_{s_F} ]   \E^\Qbbq_t[D_r \Pi^+]dt}_\text{CVA, no WWR}\nonumber  \\
&{}+  \underbrace{\int_0^T   \rho^{\text{c1}}_t    \E^\Qbbq_t[ \LGD \lambda D_{\lambda}]   \sqrt{ \Var^\Qbbq_t(  D_{s_F})  \Var^\Qbbq_t(D_r  \Pi^+)} dt}_\text{CVA\ WWR funding-exposure} \label{e:acva2}\\
&{} +   \underbrace{\int_0^T   \rho^{\text{c2}}_t \sqrt{ \Var^\Qbbq_t( \LGD \lambda D_{\lambda})  \Var^\Qbbq_t(  D_{s_F} D_r \Pi^+)} dt}_\text{CVA\ WWR default-funding+exposure}\nonumber \\
\FVA 
=&  \underbrace{\int_0^T  \E^\Qbbq_t[ D_{\lambda}]  \E^\Qbbq_t[s_F D_{s_F} ]   \E^\Qbbq_t[D_r  \Pi]dt}_\text{FVA, no WWR}\nonumber \\
&{}+  \underbrace{\int_0^T   \rho^{\text{f1}}_t    \E^\Qbbq_t[ D_{\lambda}]   \sqrt{ \Var^\Qbbq_t( s_F D_{s_F})  \Var^\Qbbq_t( D_r \Pi^)} dt}_\text{FVA\ WWR funding-exposure}  \label{e:afva2} \\
&{} +   \underbrace{\int_0^T   \rho^{\text{f2}}_t \sqrt{ \Var^\Qbbq_t( D_{\lambda})  \Var^\Qbbq_t( s_F D_{s_F} D_r \Pi^)} dt}_\text{FVA\ WWR default-funding+exposure} \nonumber
\end{align}
where
\begin{align}
 \text{Var}_t( D_{s_F} (D_{r} \Pi^+) ) =&  \rho_t^{\text{c2.1}}\sd{D_{s_F}^2}\sd{ (D_{r} \Pi^+)^2} + \Ebb[D_{s_F}^2]\Ebb[ (D_{r} \Pi^+)^2] \nonumber \\
                   &{} - \left(  \rho_t^{c1} \sd{D_{s_F}} \sd{ D_{r} \Pi^+}  +  \Ebb[D_{s_F}] \Ebb[ D_{r} \Pi^+]        \right)^2  \label{e:var2c}  \\
 \text{Var}_t( (s_F D_{s_F}) (D_{r} \Pi) ) =&  \rho_t^{f2.1}\sd{(s_F D_{s_F})^2}\sd{ (D_{r} \Pi)^2} + \Ebb[(s_F D_{s_F})^2]\Ebb[ (D_{r} \Pi)^2] \nonumber \\
                   &{} - \left(  \rho_t^{f1} \sd{s_F D_{s_F}} \sd{ D_{r} \Pi}  +  \Ebb[s_F D_{s_F}] \Ebb[ D_{r} \Pi]        \right)^2  \label{e:var2f}
\end{align}
We have suppressed most of the time indices for clarity.  Note that each term within an expectation or variance operator becomes a deterministic term structure after the expectation or variance is take.

The terminal correlation term structures used for pricing/accounting CVA are:
\begin{itemize}
\item $\rho^{\text{c1}}_t$:  $D_{s_F}$ versus $D_r  \Pi^+$
\item $\rho^{\text{c2}}_t$:  $\LGD \lambda D_{\lambda}$ versus $D_{s_F} D_r \Pi^+$
\item $\rho^{\text{c2.1}}_t$:  $D_{s_F}^2$ versus $ (D_{r} \Pi^+)^2$
\end{itemize}
These are counterparty specific.

The terminal correlation term structures used for pricing/accounting FVA are:
\begin{itemize}
\item $\rho^{\text{f1}}_t$:  $s_F D_{s_F}$ versus $D_r  \Pi$
\item $\rho^{\text{f2}}_t$:  $D_{\lambda}$ versus $s_F D_{s_F} D_r \Pi$
\item $\rho^{\text{f2.1}}_t$:  $(s_F D_{s_F})^2$ versus $ (D_{r} \Pi)^2$
\end{itemize}
These are counterparty specific.

Note that $\rho^{\text{c1}}_t$ may be quite close to $\rho^{\text{c2.1}}_t$ when $D_{s_F}$ is always non-negative.  Similarly for $\rho^{\text{f1}}_t$ and $\rho^{\text{f2.1}}_t$.  This is a property of ``reasonably behaved'' random variables with positive support, which can be verified empirically.

Note that the variances we require after using Equation \ref{e:var2c}, \ref{e:var2f} either involve credit, funding spread, or exposure.  We have arranged that no variances involving cross terms are required so these  term structures of terminal volatility can be estimated independently:
\begin{align}
\text{From\ simulation}:&\ 
\{
\Var^\Qbbq_t(D_r  \Pi^+),\ 
\Var^\Qbbq_t(D_r  \Pi), \nonumber \\  
&\ \ \ \Var^\Qbbq_t{ (D_{r} \Pi^+)^2}, \
\Var^\Qbbq_t{ (D_{r} \Pi)^2} \label{e:varS}
\}
\\
\text{From\ historical}:&\ 
\{
\Var^\Qbbq_t(  D_{s_F}), \  
\Var^\Qbbq_t( \LGD \lambda D_{\lambda}), \ 
\Var^\Qbbq_t( s_F  D_{s_F}), \nonumber \\
&\ \ \ \Var^\Qbbq_t( D_{\lambda}), \   
\Var^\Qbbq_t(  D_{s_F}^2 ), \ 
\Var^\Qbbq_t{(s_F D_{s_F})^2}  \label{e:varH}
\}
\end{align}

\subsubsection{Parameter Estimation}
\label{s:estCFva}

This follows the same pattern as for regulatory CVA.  Expectations are market-implied as are exposure variances.  Default variances, funding variances and term structures of terminal correlation are historically calibrated.

\subsubsection{Properties}

\begin{itemize}
\item Equations \ref{e:acva2} and \ref{e:afva2} for CVA and FVA respectively, are exact: this is a trivial property since we are only re-writing the original equations \ref{e:acva1} and \ref{e:afva1}.  No assumptions are required on exposure underlyings, default, or correlation beyond the existence of second moment for underlyings and squares of underlyings.
\item There are three term structures of terminal correlation that determine the price of WWR for CVA, and another three for WWR of FVA.
\item The terminal correlation term structures are deterministic.  
\item The model cleanly separates the items that may be market implied (the expectations and exposure variances) from those that need historical estimation (terminal correlations and variances of default and funding) as before.  
\item Having  two WWR terms for accounting CVA and FVA is a first order property that is simply the result of having three underlyings: exposure; default; and funding.
\end{itemize}

In the following numerical examples we will show in detail  the historical estimation of variances and correlations.  

\section{Numerical examples}

We first describe the setup, then calibration, and then numerical results.  The setup is an example and its elements are not prescriptive, for example funding costs might be idiosyncratic, or derived from Totem.

\subsection{Setup}

We consider WWR for the experimental setup of a bank trading with an uncollaterlized counterparty as follows
\begin{description}
\item[asof] 2019-11-01
\item[trades] vanilla EUR interest rate swaps with maturities of 5, 10, 20, and 30 years.  Receive float and receive fixed.
\item[counterparty CDS] Eur iTraxx crossover
\item[bank funding spread] Eur iTraxx Senior Financials
\item[correlation calibration period] 2008 -- 2012, for all correlations
\item[counterparty CDS volatility calibration period] one year up to asof date
\item[swaption volatiliy] ATM Normal volatilities for both historical period and valuation date.
\end{description}
Asof is a recent date.  The trades maturities cover a normal range of lengths for counterparty IRS trades.   The counterparty CDS spread is a choice that represents counterparties that may have significant CVA and so are of interest for WWR.  We use Eur iTraxx Senior Financials to represent the funding spread over riskless for the bank as an observable and related spread.  The correlation calibration period is inspired by MAR50 in that it covers five years and includes a period of stress for interest rate trades.  There are no liquid single name CDS options, and liquid index CDS options have maximum expiry of 9 months, so we use historical calibration for CDS volatility, picking a recent period, i.e. last one year.

Since we are working with vanilla IRS we have no need of simulation model for CVA and FVA, we simply use the sum-of-risky-swaptions approach.  On the valuation date we do need the volatilities of the exposure levels and use the Normal distribution from the market data to obtain this.  We use low-discrepancy integration for the exposure and similar volatilities for efficiency, but there is effectively no numerical noise.

\subsection{Calibration}

The calibration process follows the outline in Sections \ref{s:estCva} and \ref{s:estCFva}.

\subsubsection{Common features}

The portfolio is kept constant as it is evaluated against the different market data from different dates, similarly with how VaR is calculated.  Default probabilities are calculated asof their market data dates because CDS maturities are standardized on IMM dates.  Fixings are taken from the valuation date and inserted into the historical market data files.

\subsubsection{Correlations}

An outline of the correlation calibration process is shown in Figure \ref{f:cal}.  We prove in Appendix 2 that this method provides the correlation of the underlyings even though we calculate the correlation between expectations of the underlyings.   Here we describe the process for regulatory CVA.  The process for accounting CVA and FVA is similar.  In the Figure the objective is to get the correlation between the default probability (PD) between $\tau$ to $\tau+dt$ and the corresponding positive expected exposure (EE).  

For a counterparty, the market data for each date in the calibration period provides one exposure profile, and one counterparty default probability profile.  These profiles are shown at the top of  Figure \ref{f:cal}.  So for a five year calibration period we have 252 x 5 = 1260 profiles of exposure and 1260 profiles of default probability, one for each market data date.  Each date is a conditional sample of the expected exposure and expected default probability.  The conditioning is on the market data date.

Considering a single simulation interval, say $\tau$ to $\tau+dt$ we can obtain an expected exposure from each of the 1260 profiles.  We can also obtain 1260 corresponding samples of the expected default probability from  $\tau$ to $\tau+dt$.  We plot these 1260 pairs of values (expected exposure, probability of default) to obtain the bottom plot of \ref{f:cal}.  The example is with three market data dates.  On each of the three dates shown we compute the EE and PD against time (aka portfolio maturity).  We can then calculate the correlation between these 1260 pairs of values to obtain $\rho(\tau, dt)$ for the simulation interval $\tau$ to $\tau+dt$ from time $\tau$ to $\tau+dt$.

We repeat the process in Figure \ref{f:cal} for every time (aka forward maturity) $\tau$ over the life of the portfolio with the counterparty (``Time'' in the Figure).  This set of correlations versus Time provides the correlation term structure that goes into the WWR calculation. 

We use correlations between sample expected exposure and sample expected default probability in place of correlations between  exposure and  default probability.  In Appendix 2 we prove that these two correlations are the same.

\subsubsection{Expected default probability standard deviations}

The term structure of expected default probability standard deviation, including LGD changes, is calculated historically, in a similar way to the correlation term structure except that the most recent one year period is used, as it would be for pricing.

\subsubsection{Expected exposure standard deviations}

Expected exposure standard deviations are calculated from the usual portfolio simulation.  That is, we take the exposure at $\tau$ from each of the simulation paths and calculate a standard deviation from these.  This gives the standard deviation of exposure for $\tau$.  We repeat this for every simulation time (aka forward maturity) $\tau$ over the life of the portfolio with the counterparty to obtain the term structure of exposure standard deviation.

\subsubsection{Standard deviations for accounting CVA and FVA}

The standard deviations needed for WWR for accounting CVA and FVA are listed in Equation \ref{e:varS} where they can be estimated from usual XVA simulation and in  Equation \ref{e:varH} where they must be estimated from historical data.  These copy the processes described above.

\begin{figure}
\centering
\includegraphics[width=\textwidth,trim=0 0 0 0,clip]{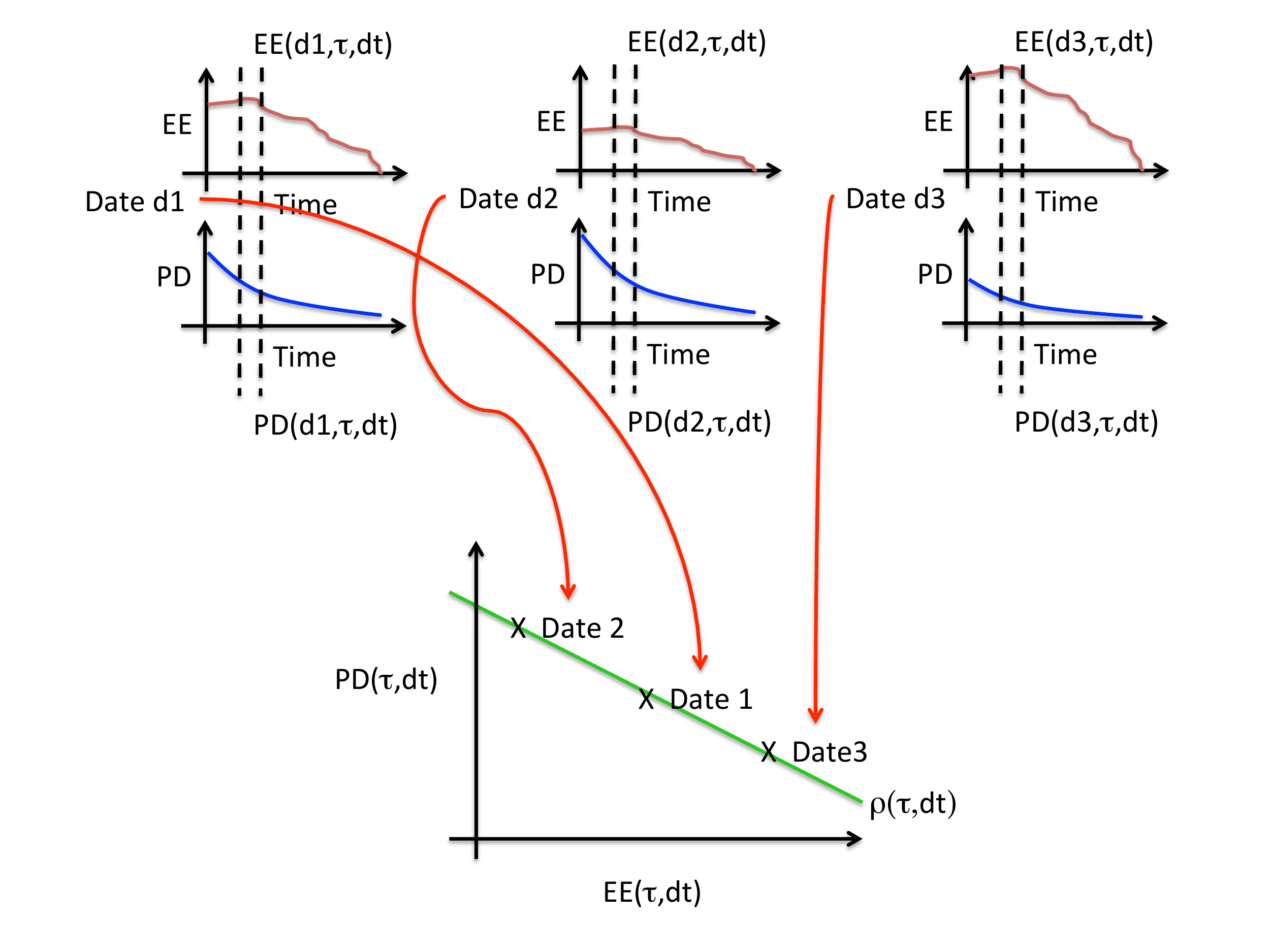}
\caption{Outline of correlation calibration process.  For details see text.}
\label{f:cal}
\end{figure}

\subsection{Results}

\begin{figure}
\centering
\includegraphics[trim=0 0 0 0,clip,width=0.45\textwidth]{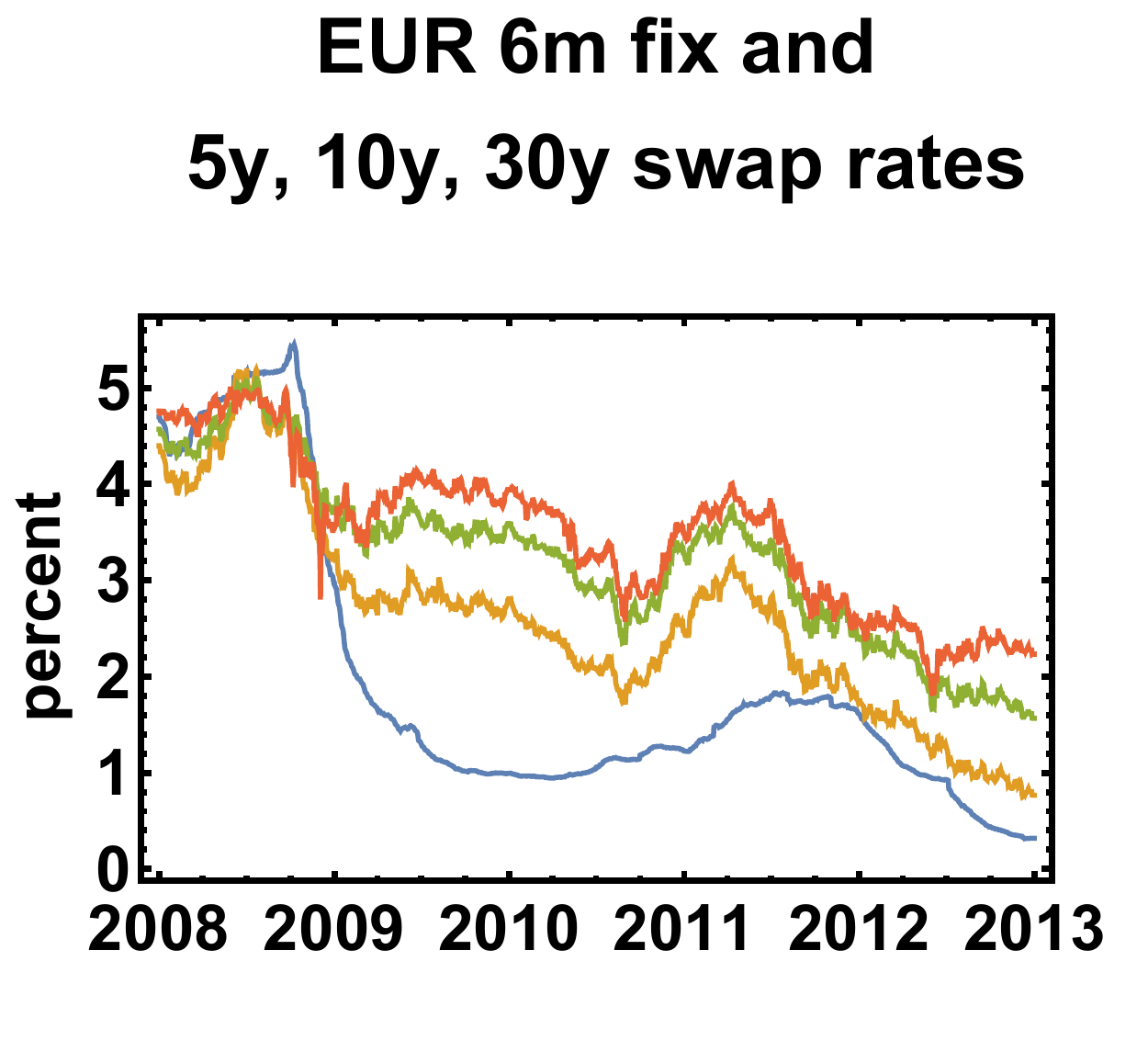}
\includegraphics[trim=0 0 0 0,clip,width=0.45\textwidth]{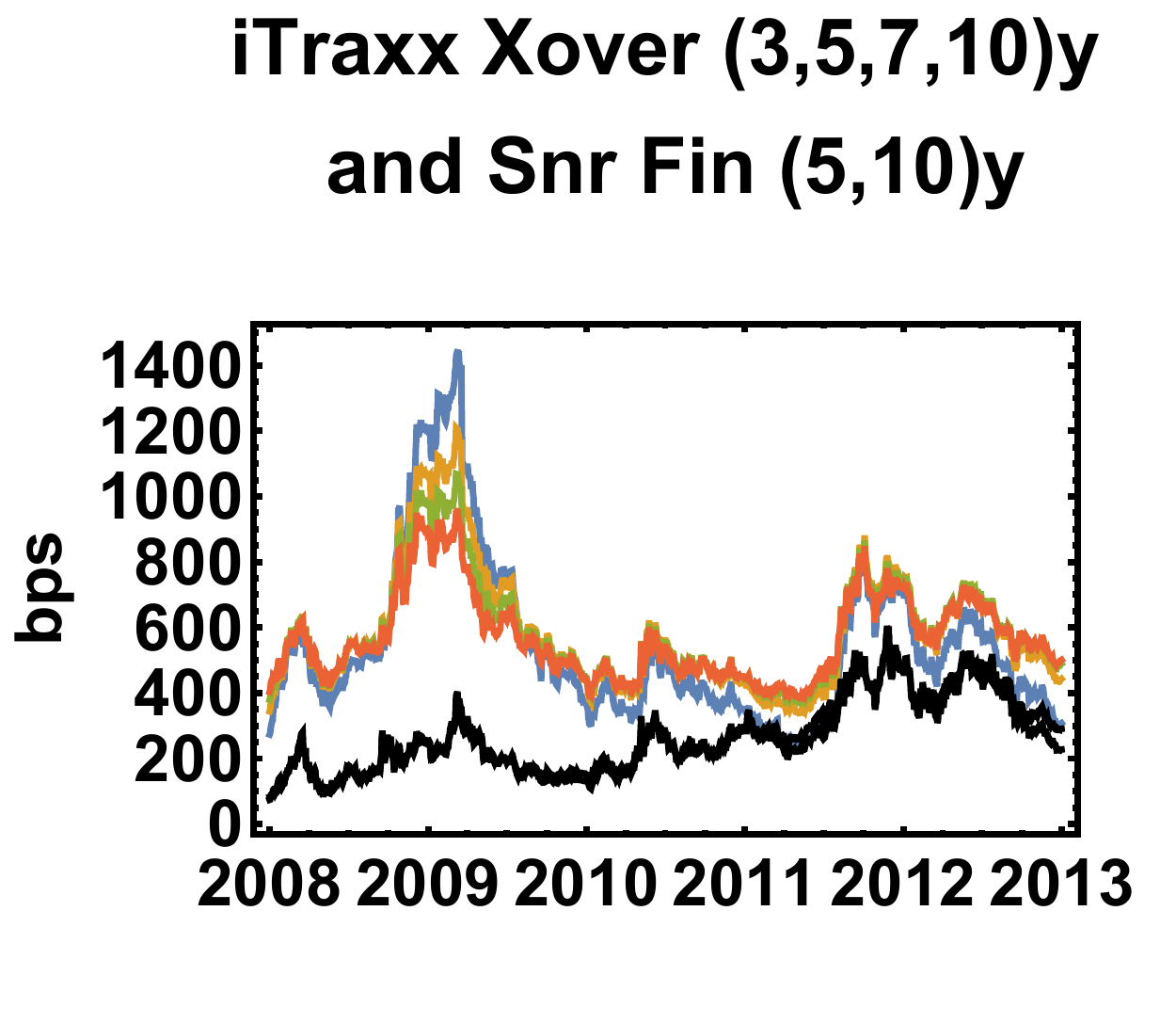}
\caption{Jan 2008 -- Dec 2012 market data for correlation calibration. LHS interest rates.  RHS CDS spreads: iTraxx crossover are in colour; Senior Financials are in black.}
\label{f:ircds}
\end{figure}

We show a sample of the input data as well as all the WWR numbers for the trades considered.  Figure \ref{f:ircds} shows the history of EUR 6m fixings and \{5y, 10y, 30y\} swap rates together with Eur iTraxx Xover and Eur Senior Financials.  

\subsubsection{Regulatory CVA}

\begin{figure}
\centering
\includegraphics[trim=0 0 0 0,clip,width=0.45\textwidth]{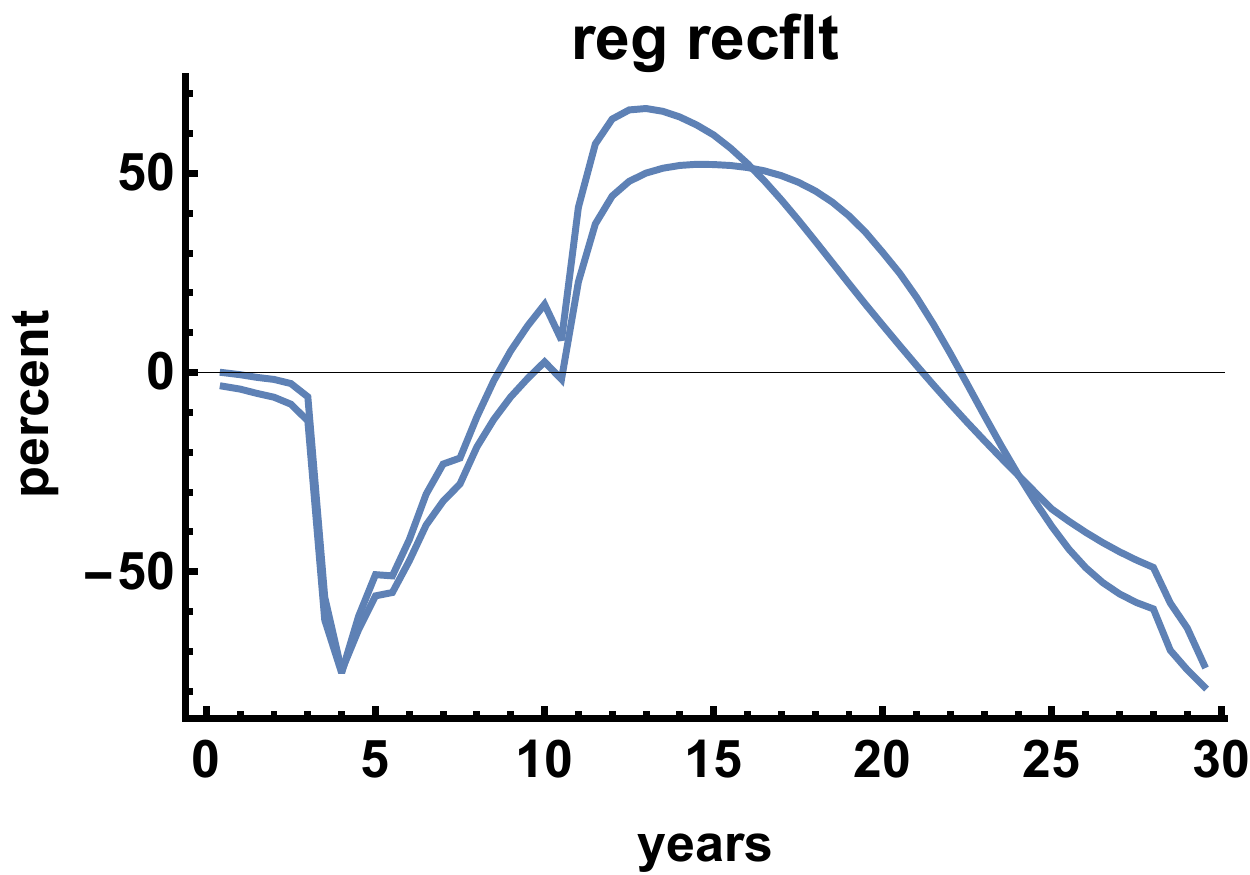}
\includegraphics[trim=0 0 0 0,clip,width=0.45\textwidth]{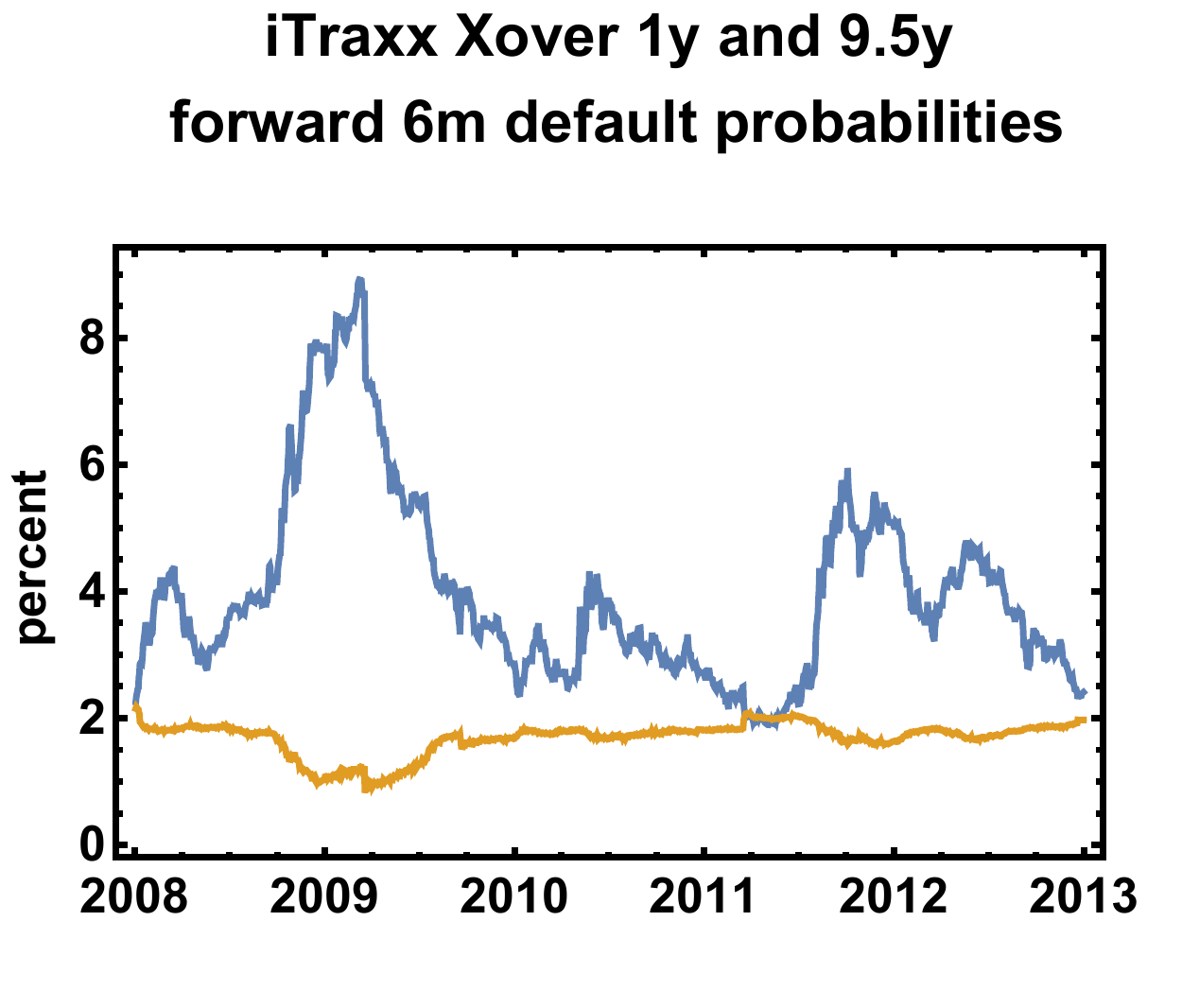}
\includegraphics[trim=0 0 0 0,clip,width=0.45\textwidth]{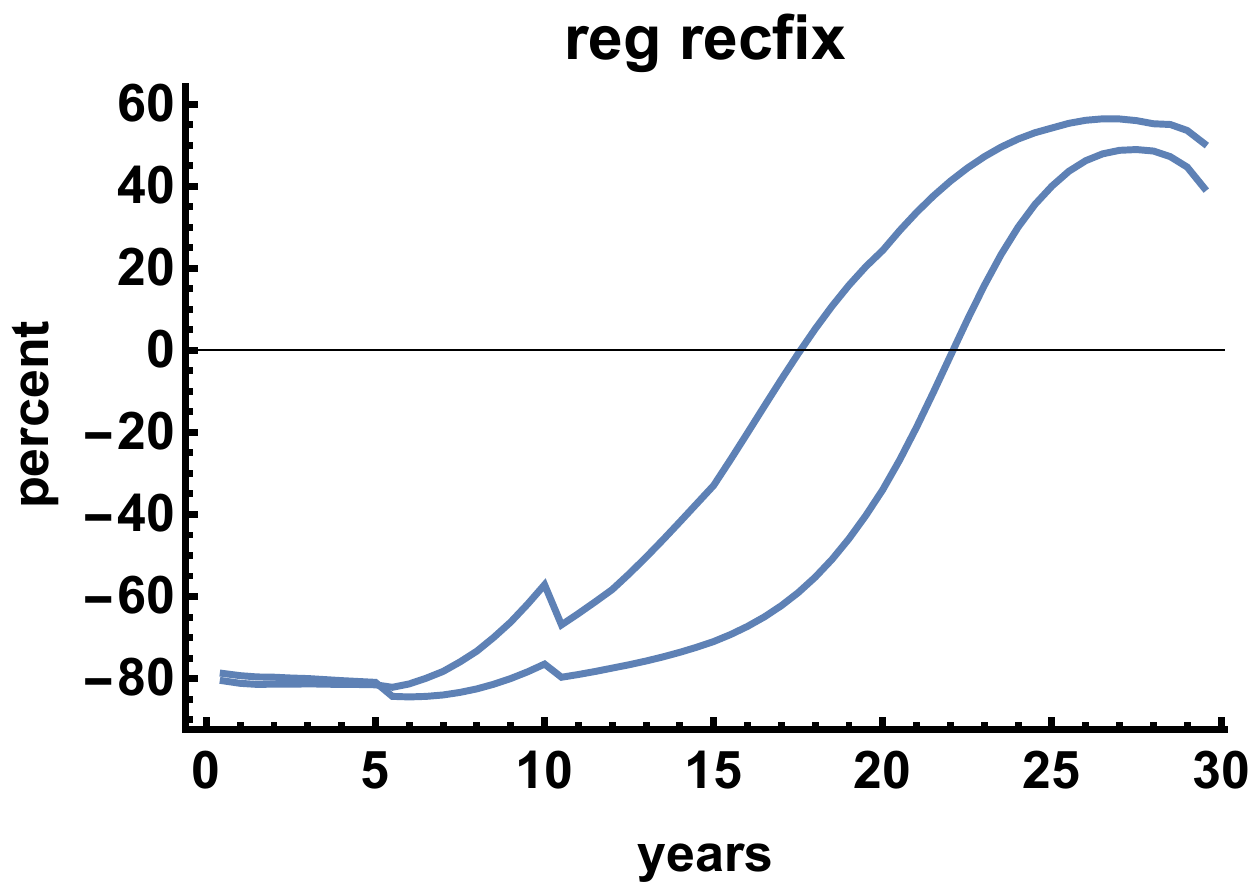}
\includegraphics[trim=0 0 0 0,clip,width=0.45\textwidth]{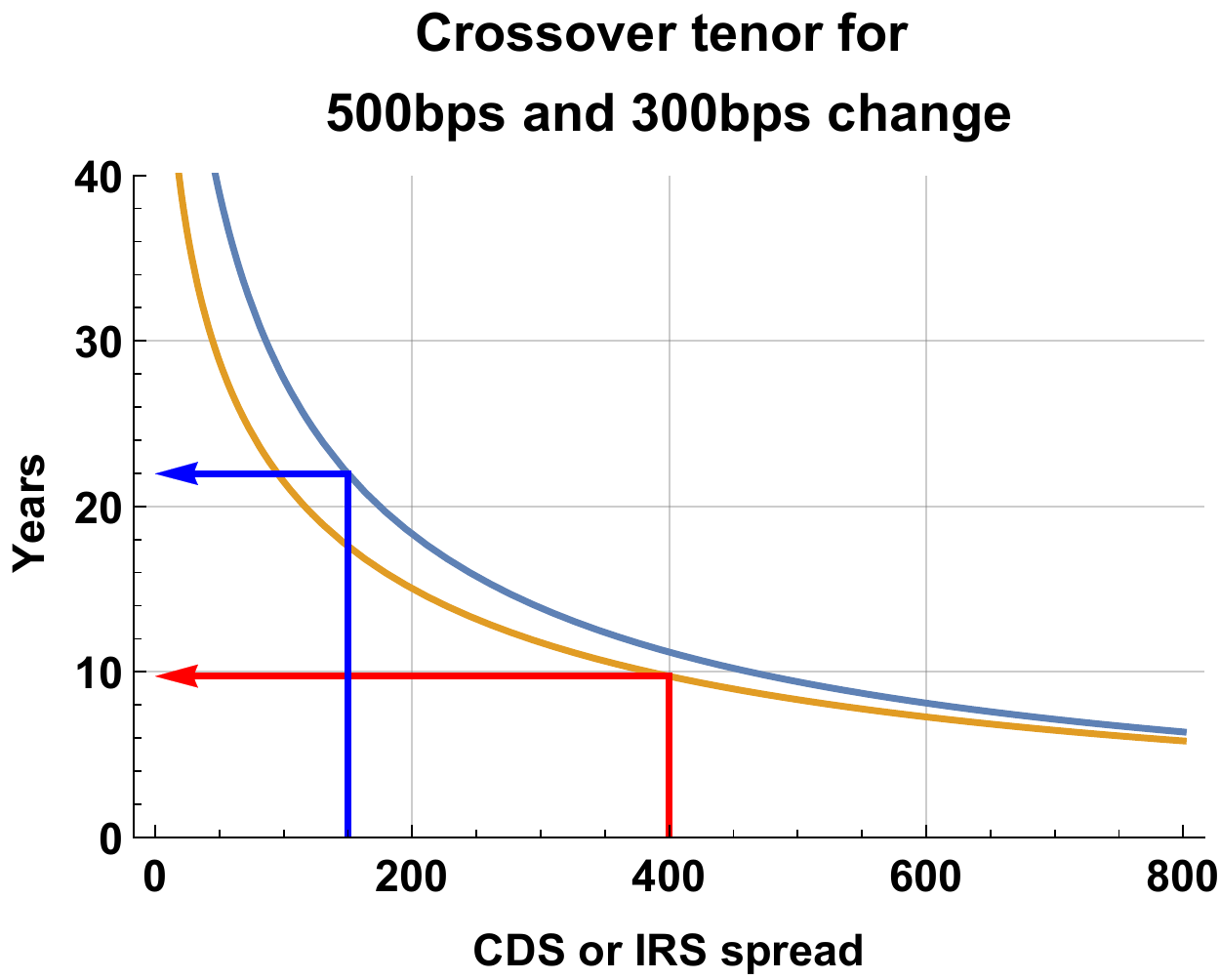}
\caption{
Top Left: regulatory CVA exposure-default correlation term structure for 30Y receive float IRS, different curves are different strikes (-25bps and 2\%).  
Top Right: iTraxx Xover 6m forward default probabilities.
Bottom Left: regulatory CVA exposure-default correlation term structure for 30Y receive fix IRS, different curves are different strikes (-25bps and 2\%).
Bottom Right:  crossover tenors for 500bps (brown) and 300bps (blue) changes in CDS or IRS levels.  Arrows indicate predicted crossovers for receive IRS: compare with Top Left. For receive fixed IRS (Bottom Left) the rates move dominates the credit move.
}
\label{f:edcorr}
\end{figure}

Figure \ref{f:edcorr} shows the exposure-default correlation term structure for 30Y ATM receive float IRS in the Top Left plot.  The wave-like shape is superficially unexpected but can be understood from the plot of forward 6m default probabilities (Top Right).  Observe that when the CDS spreads increase in 2009 the 1y forward default probability increases significantly, but the 9.5y forward default probability decreases.  Since the area under the forward default probability curve must integrate to unity, an increase in one region must be balanced by a decrease in another region.  

 We can compute the crossover tenor between increase and decrease of default probabilities by equating the instantaneous forward default probabilities:
\begin{align}
\lambda_1 e^{-\lambda_1 t} =& \lambda_2 e^{-\lambda_2 t} \nonumber\\
\text{crossover\ point\ } t =& \frac{\log(\lambda_1/\lambda_2)}{\lambda_1-\lambda_2}  \label{e:xover}
\end{align}
Suppose we have a flat CDS curve and we increase it by 500bps, which is similar to the smallest increase for the iTraxx Xover during the crisis, this then gives the brown curve on the Bottom Right of Figure \ref{f:edcorr}.

The second crossover tenor in the correlation term structure is also due to curve movements over 2008--2012.  Consider the interest rates as though they were CDS, simply because an IRS pays a discounted rate which has a similar form to the instantaneous default probability $\lambda \exp(-t\lambda)$.  We can then apply the same logic as above using the blue curve in Bottom Right plot of  Figure \ref{f:edcorr} considering the decrease in EUR rates from around 4.5\%\ to around 1.5\%\ over the period.  This gives a predicted crossover around 20y which is similar to that observed in both receive float IRS (Top Left) and receive fixed IRS (Bottom Left).  For the receive fixed IRS the interest rate move dominates the credit move so there is no early crossover in the exposure-default correlation term structure.

Shorter IRS, and IRS with different strikes have similar patterns to the correlation term structures shown, but truncated in the case of shorter IRS.

\begin{table}[ht]
\begin{adjustwidth}{-3.3cm}{-3.3cm}
\begin{centering}
\begin{tabular}{rrrrrrrrrr}
T     & K       & CVA Indep & WW & WW+Crisis & T     & K       & CVA Indep & WW & WW+Crisis \\
years & percent & RecFlt    & RecFlt & RecFlt        & years & percent & RecFix    & RecFix & RecFix        \\ \hline
5     & -0.25   & 5         & -0.1   & 0.            & 5     & -0.25   & 3         & -0.5   & 0.            \\
5     & 0       & 2         & -0.1   & 0.            & 5     & 0       & 8         & -0.7   & 0.            \\
5     & 0.5     & 0         & 0.     & 0.            & 5     & 0.5     & 19        & -0.9   & 0.            \\
5     & 1.      & 0         & 0.     & 0.            & 5     & 1.      & 33        & -1.    & 0.            \\
5     & 2.      & 0         & 0.     & 0.            & 5     & 2.      & 60        & -1.    & 0.            \\
10    & -0.25   & 80        & -2.7   & -0.2          & 10    & -0.25   & 18        & -3.1   & -0.2          \\
10    & 0       & 58        & -2.4   & -0.2          & 10    & 0       & 27        & -3.9   & -0.2          \\
10    & 0.5     & 30        & -1.9   & -0.2          & 10    & 0.5     & 61        & -5.4   & -0.3          \\
10    & 1.      & 14        & -1.3   & -0.1          & 10    & 1.      & 106       & -6.6   & -0.3          \\
10    & 2.      & 2         & -0.5   & -0.1          & 10    & 2.      & 218       & -7.6   & -0.4          \\
20    & -0.25   & 480       & -10.4  & 8.2           & 20    & -0.25   & 77        & -9.1   & -8.           \\
20    & 0       & 397       & -9.7   & 7.6           & 20    & 0       & 103       & -11.3  & -9.4          \\
20    & 0.5     & 256       & -8.1   & 6.2           & 20    & 0.5     & 182       & -16.1  & -12.3         \\
20    & 1.      & 160       & -6.3   & 5.            & 20    & 1.      & 305       & -20.7  & -15.          \\
20    & 2.      & 55        & -3.    & 2.9           & 20    & 2.      & 639       & -26.5  & -19.3         \\
30    & -0.25   & 908       & -12.1  & 28.6          & 30    & -0.25   & 193       & -13.5  & 2.5           \\
30    & 0       & 756       & -11.5  & 28.8          & 30    & 0       & 249       & -17.3  & -3.8          \\
30    & 0.5     & 497       & -9.7   & 26.9          & 30    & 0.5     & 406       & -25.6  & -17.8         \\
30    & 1.      & 317       & -7.3   & 23.4          & 30    & 1.      & 641       & -33.4  & -32.          \\
30    & 2.      & 117       & -3.1   & 15.           & 30    & 2.      & 1272      & -42.9  & -56.9     \\ \hline
\end{tabular}
\caption{Regulatory CVA for Receive-Floating (RecFlt) and Receive-Fixed (RecFix) vanilla EUR IRS for a range of tenors (T) and strikes (K).  We show WWR for CVA under current conditions (CVA+WW) and under conditions with increased CDS volatility taken from 2008 as WW+Crisis.  Note that strikes are absolute (not relative to ATM).  Units are bps of notional.}
\label{t:reg}
\end{centering}
\end{adjustwidth}
\end{table}

Table \ref{t:reg} shows independent accounting CVA and the regulatory WW component in bps of notional for a range of tenors and strikes.  We see that WWR for receive-float IRS (uncommon case) can be up to 25\%\ of independent CVA but is typically a few percent.  For receive-fixed IRS (usual case with clients) the WWR is also a few percent.  In both cases crisis levels of CDS volatility reduce WWR CVA.

\FloatBarrier
\subsubsection{Accounting CVA and FVA}

We now include funding, modeled using the iTraxx Senior Financials CDS index points shown in black in Figure \ref{f:ircds} RHS.

\begin{figure}
\centering
\includegraphics[trim=0 0 0 0,clip,width=0.45\textwidth]{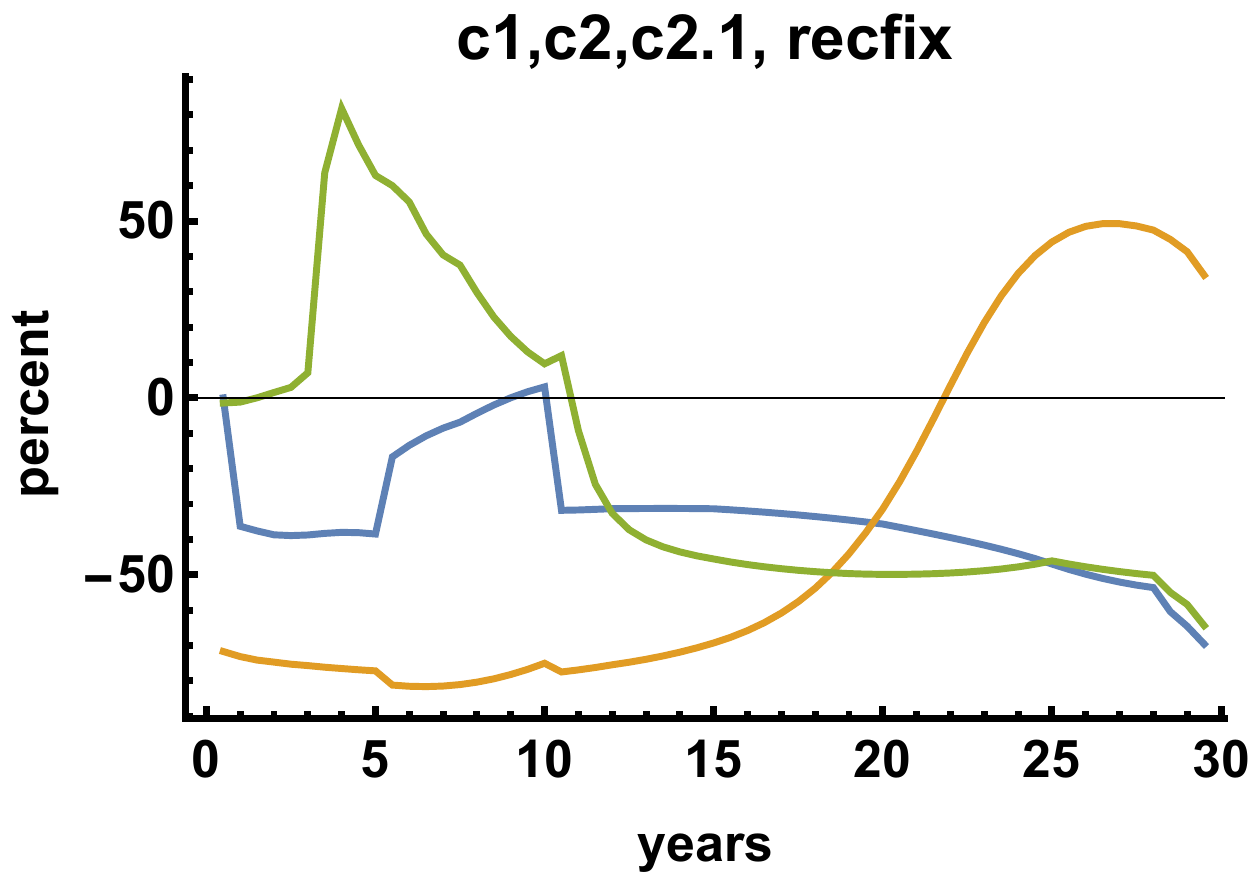}
\includegraphics[trim=0 0 0 0,clip,width=0.45\textwidth]{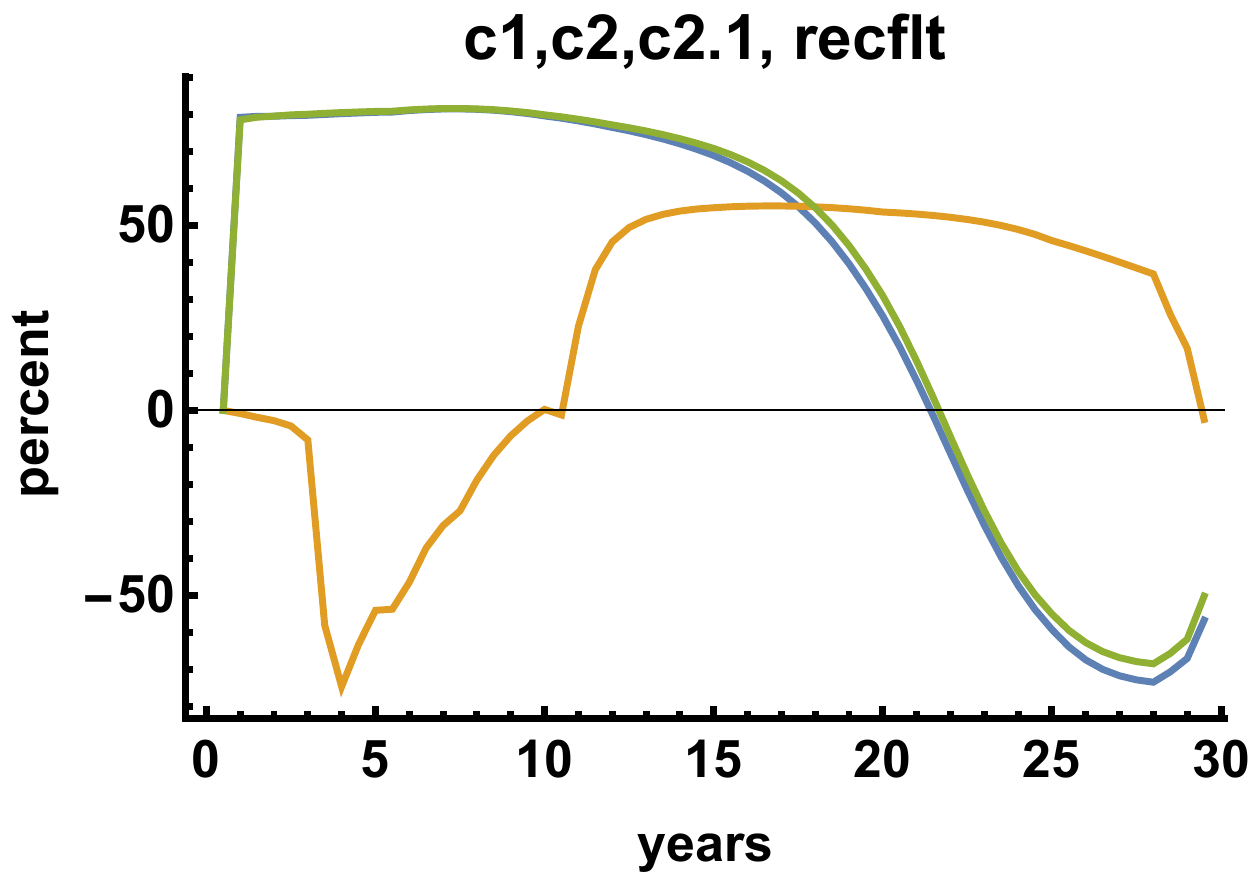}
\includegraphics[trim=0 0 0 0,clip,width=0.45\textwidth]{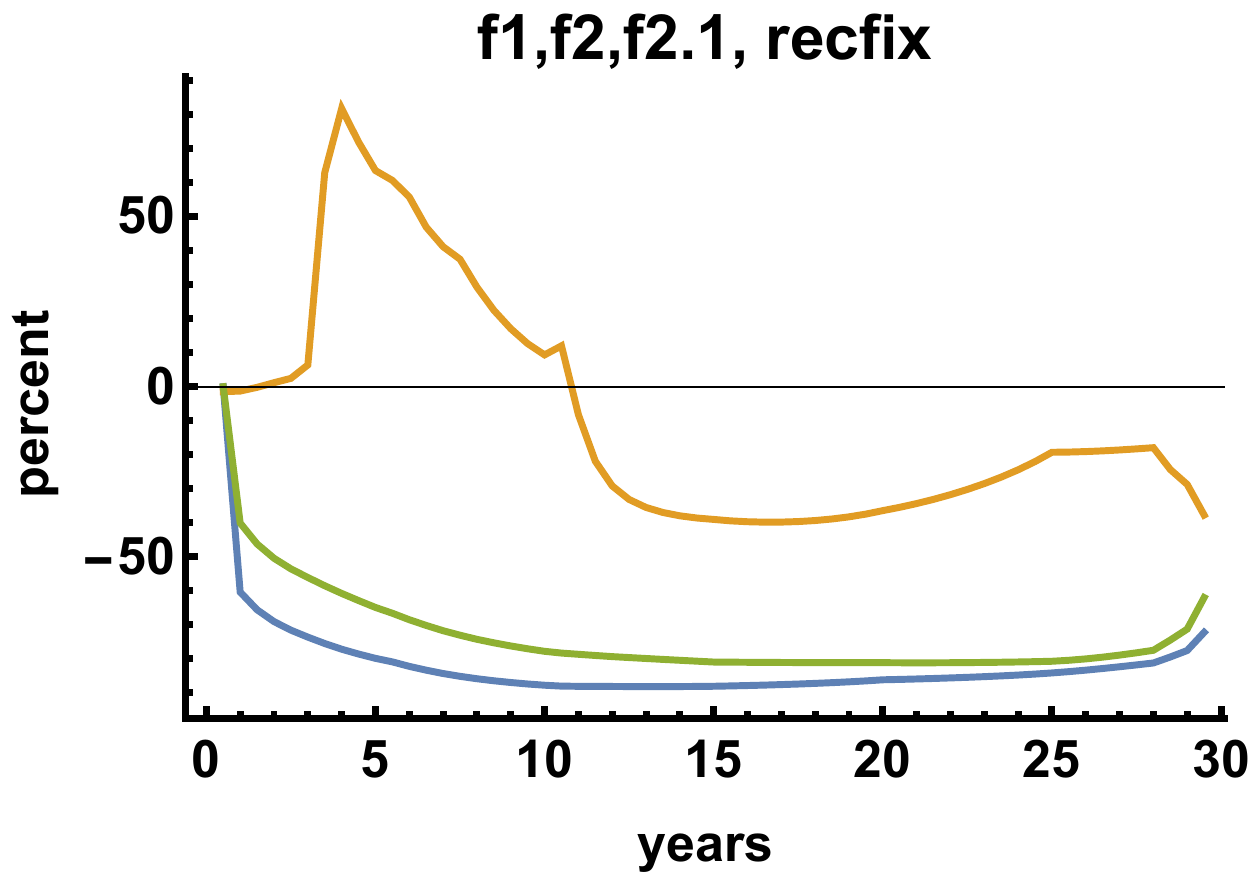}
\includegraphics[trim=0 0 0 0,clip,width=0.45\textwidth]{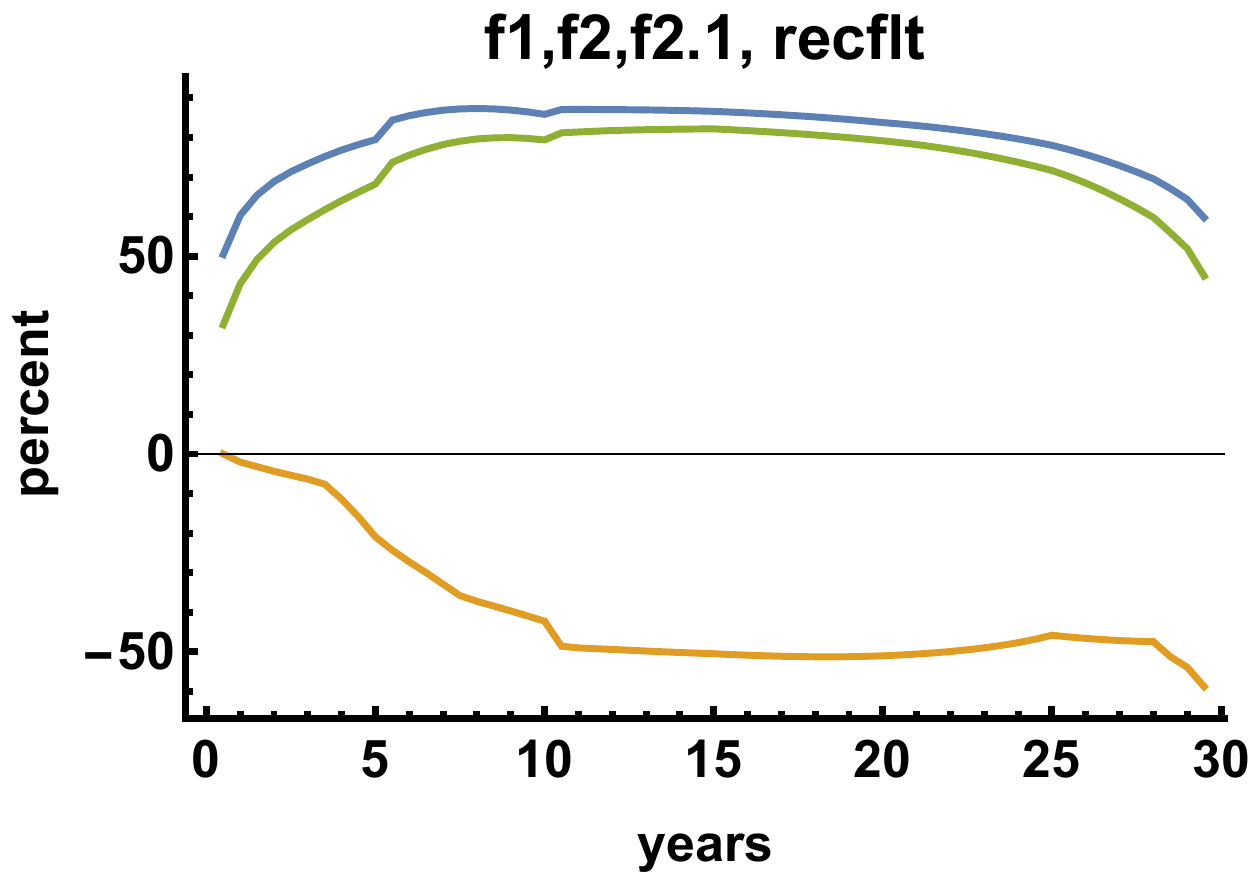}
\caption{
Top Left:  accounting CVA \{c1,c2,c2.1\} correlations for 30y receive fix IRS.  
Top Right: receive float IRS.
Bottom Left:   FVA \{f1,f2,f2.1\} correlations for receive fix IRS.
Bottom Right: receive float IRS.
Blue and Green curves are *1 and *2.1 correlations, these have the same underlyings but squared in the second case. *2 are the yellow curves.
}
\label{f:acccorr}
\end{figure}

Figure \ref{f:acccorr} shows the correlation term structures for accounting CVA \{c1,c2,c2.1\} and FVA \{f1,f2,f2.1\} for receive float IRS and receive fix IRS.  Overall the correlation patterns are a combination of the previous regulatory case, and a new behavior when there are no changes of sign of the correlation term structures.  These behaviors are a reflection of the more complex nature of the WWR terms and that FVA is not option-like (uses net exposure), unlike CVA (positive exposure only).

Table \ref{t:acc1} gives the CVA and FVA details for the range of vanilla EUR receive float IRS.  Table \ref{t:acc1} provides the details for receive fixed IRS.

We can first note that the second WWR term for both CVA and FVA is negligible.  This second WWR term is a mixture of all three factors mixed in different ways by the variance terms so, net, there is no directional contribution.  It is possible that this observation will also hold for other instruments.

In absolute terms, and compared to the regulatory view, accounting CVA WWR is small for receive fixed IRS, mostly below or of the order of a couple of percent.  In contrast accounting CVA WWR for receive float IRS can be above 10\%\ although it is also generally below a couple of percent.

FVA WWR is significant as it is often more than ten percent and can reach more than fifty percent in some cases.  The greater magnitude of FVA WWR compared to accounting CVA  WWR is mostly because it is not option-like so has simpler correlation behavior in general.

\begin{table}[ht]
\begin{adjustwidth}{-3.3cm}{-3.3cm}
\begin{centering}
\begin{tabular}{rrrrrrrrrr}
T     & K       & CVA Indep & WW1 &WW2 & FVA Indep & WW1 & WW2 & CVA    & FVA    \\
years & percent & RecFlt    & RecFlt  & RecFlt  & RecFlt    & RecFlt  & RecFlt  & RecFlt & RecFlt \\ \hline
5     & -0.25   & 5         & 0       & 0       & 1         & 0       & 0       & 5      & 0.8    \\
5     & 0       & 2         & 0       & 0       & -3        & 0       & 0       & 2      & -2     \\
5     & 0.5     & 0         & 0.1     & -0.0001 & -9        & 1       & 0       & 0      & -7.5   \\
5     & 1       & 0         & 0.1     & -0.0002 & -15       & 2       & 0       & 0      & -12.7  \\
5     & 2       & 0         & 0.2     & -0.0003 & -27       & 4       & 0       & 0      & -23    \\
10    & -0.25   & 73        & 0.1     & -0.0002 & 25        & 1       & -0.0001 & 73     & 26     \\
10    & 0       & 55        & 0.2     & -0.0002 & 12        & 2       & -0.0001 & 55     & 14     \\
10    & 0.5     & 28        & 0.4     & -0.0003 & -13       & 4       & -0.0001 & 29     & -9.4   \\
10    & 1       & 13        & 0.7     & -0.0006 & -38       & 6       & -0.0001 & 14     & -32.1  \\
10    & 2       & 2         & 1.4     & -0.0013 & -89       & 13      & -0.0001 & 3      & -75.7  \\
20    & -0.25   & 425       & 1.1     & -0.0004 & 166       & 4       & 0.0003  & 426    & 170    \\
20    & 0       & 352       & 1.4     & -0.0004 & 121       & 5       & 0.0003  & 354    & 126    \\
20    & 0.5     & 230       & 2.4     & -0.0005 & 30        & 9       & 0.0003  & 232    & 39.6   \\
20    & 1       & 142       & 3.7     & -0.0007 & -60       & 16      & 0.0003  & 146    & -44.5  \\
20    & 2       & 48        & 7.2     & -0.0019 & -241      & 35      & 0.0003  & 55     & -206   \\
30    & -0.25   & 783       & 1.1     & -0.0004 & 291       & 8       & 0.002   & 784    & 298.9  \\
30    & 0       & 652       & 1.9     & -0.0004 & 206       & 11      & 0.002   & 654    & 217.2  \\
30    & 0.5     & 430       & 4       & -0.0005 & 37        & 19      & 0.002   & 434    & 56.2   \\
30    & 1       & 270       & 6.9     & -0.0007 & -131      & 31      & 0.0019  & 277    & -100.5 \\
30    & 2       & 96        & 14.6    & -0.0018 & -469      & 68      & 0.0019  & 111    & -401.3 \\ \hline
\end{tabular}
\caption{Accounting CVA and FVA for Receive-Floating (RecFlt)  vanilla EUR IRS for a range of tenors (T) and strikes (K).   Note that strikes are absolute (not relative to ATM).  Units are bps of notional. Last two columns are total CVA and FVA.  WW1 and WW2 are the first and second WWR terms in Equations \ref{e:acva2} and  \ref{e:afva2}.}
\label{t:acc1}
\end{centering}
\end{adjustwidth}
\end{table}

\begin{table}[ht]
\begin{adjustwidth}{-3.3cm}{-3.3cm}
\begin{centering}
\begin{tabular}{rrrrrrrrrr}
T     & K       & CVA Indep & WW1 & WW2 & FVA Indep &WW1 &WW2 & CVA    & FVA    \\
years & percent & RecFix    & RecFix  & RecFix  & RecFix    & RecFix  & RecFix  & RecFix & RecFix \\ \hline
5     & -0.25   & 3         & 0       & -0.0001 & -1        & 0       & 0       & 3      & -0.4   \\
5     & 0       & 7         & 0       & -0.0001 & 3         & 0       & 0       & 7      & 2.3    \\
5     & 0.5     & 16        & 0       & -0.0003 & 9         & -1      & 0       & 16     & 7.7    \\
5     & 1       & 27        & -0.1    & -0.0005 & 15        & -2      & 0       & 27     & 13.1   \\
5     & 2       & 51        & -0.1    & -0.0009 & 27        & -3      & 0       & 50     & 23.8   \\
10    & -0.25   & 17        & 0.2     & -0.0006 & -25       & 3       & -0.0001 & 17     & -21.7  \\
10    & 0       & 26        & 0.1     & -0.0004 & -12       & 2       & -0.0001 & 26     & -10.6  \\
10    & 0.5     & 54        & -0.1    & -0.0007 & 13        & -2      & -0.0001 & 54     & 11.4   \\
10    & 1       & 94        & -0.3    & -0.0015 & 38        & -5      & -0.0001 & 94     & 33.4   \\
10    & 2       & 193       & -0.8    & -0.0029 & 89        & -12     & -0.0001 & 192    & 77.1   \\
20    & -0.25   & 68        & 2.1     & -0.0025 & -166      & 18      & -0.001  & 70     & -147.8 \\
20    & 0       & 92        & 1.6     & -0.0018 & -121      & 14      & -0.0011 & 94     & -107.2 \\
20    & 0.5     & 164       & 0.5     & -0.0009 & -30       & 4       & -0.0012 & 164    & -26.6  \\
20    & 1       & 269       & -0.5    & -0.002  & 60        & -7      & -0.0012 & 269    & 53.1   \\
20    & 2       & 562       & -2.3    & -0.0055 & 241       & -31     & -0.0012 & 560    & 210.9  \\
30    & -0.25   & 162       & 4.4     & -0.0033 & -291      & 32      & -0.0004 & 167    & -259.1 \\
30    & 0       & 211       & 3.1     & -0.0024 & -206      & 23      & -0.0007 & 214    & -183.2 \\
30    & 0.5     & 349       & 0.6     & -0.001  & -37       & 5       & -0.0011 & 350    & -32.8  \\
30    & 1       & 549       & -1.8    & -0.0025 & 131       & -16     & -0.0014 & 547    & 115.8  \\
30    & 2       & 1094      & -5.5    & -0.0074 & 469       & -59     & -0.0018 & 1088   & 409.5  \\ \hline
\end{tabular}
\caption{Accounting CVA and FVA for Receive-Fixed (RecFix)  vanilla EUR IRS for a range of tenors (T) and strikes (K).   Note that strikes are absolute (not relative to ATM).  Units are bps of notional.  Last two columns are total CVA and FVA.   WW1 and WW2 are the first and second WWR terms in Equations \ref{e:acva2} and  \ref{e:afva2}.}
\label{t:acc2}
\end{centering}
\end{adjustwidth}
\end{table}

\FloatBarrier
\section{Discussion and Conclusions}

The key issue with WWR within XVA is calibration: no liquid instruments exist, and specifically the volatility of default probabilities is actually unobservable.  Thus we propose an alternative definition of WWR of XVA that we have argued is both useful and appropriate: this measures  the WWR that users of XVA where credit, exposure, and funding, assumed independent will observe over time.  We further prove in Appendix 2 that the term structure of correlation we use is the same as that for the underlyings.  

The model independent WWR approach presented here is unique because it requires no additional parameterized model as is the usual focus for WWR, but is simply a re-writing of the expressions for regulatory CVA and accounting CVA and FVA, and the focus is on the calibration. Once an approach is model independent it is unique because it does not depend on a choice of models.  It does however, depend on a choice of calibration data.  Being data driven, rather and model driven, is key to our approach.  Because the approach is model independent any calibration is directly in terms of the parameters for CVA and FVA calculation themselves and so provides maximum transparency, given the update for the WWR definition.  

WWR depends on correlations of factors and here we use the elements in the CVA and FVA calculations themselves: default; exposure; funding.  Correlation calibration is usually historical and this is what we do here.  We have picked a historical period inspired by the  MAR50.10, CRE53.7, and d499 (margin for non-centrally cleared derivatives) regulations in that it encompasses a stressed period (around the end of 2008) and a longer unstressed period. Because the components of CVA and FVA are estimated from separate sources the correlation might be inconsistent with what is achievable from the distributions if the components were estimated from the same source.  Whilst this is theoretically possible we consider it unlikely in practice.

The model independent WWR approach presented here is a  data driven approach, that is we estimate detailed parameters (term structures) resulting from re-writing the equations for CVA and FVA.  This provides parameter transparency in terms of the parameters driving CVA and FVA directly with details relevant to the specific portfolios considered.  In exchange it provides limited insight into the mechanism creating the observed parameters, i.e. risk factor drivers.  In contrast parameters for risk factor drivers offer limited insight into the effects on CVA and FVA for a specific portfolio.  Risk factor drivers and CVA/FVA expression  parameters offer complementary approaches to WWR. 

In use the CVA/FVA parameters of model independent approach would be recalibrated either on a schedule, say monthly, or event driven if there are significant changes to the portfolio.  Non-collateralized portfollios with clients tend to have persistence so periodic recalibration is reasonable.

XVA usually incorporates funding costs which are not part of regulations, but funding costs are of immediate interest to accounting (aka pricing).  We were able to extend the model independent approach naturally to include the third factor (funding) with the same degree of transparency as for regulatory CVA.  The increase in complexity, three correlations rather than one and two WWR terms rather than one, is also expected given we now have three factors rather than two.  However, our results indicate that the second WWR term may be negligible thus reducing complexity.

We observe that structural shifts during the calibration period (2008-2012) are reflected in the correlation term structures, notably in the sign changes of correlations.  The effect on CVA and FVA is a complex interplay between the portfolio (IRS here), credit and funding.  Because our CVA and FVA inputs are directly in terms of CVA and FVA we can both observe and numerically/analytically explain these effects (see Figure \ref{f:edcorr} and Equation \ref{e:xover}).

Our results suggest that WWR in CVA strongly depends on the instrument, but may generally be low, i.e. of the order of a couple of percent.  In contrast, WWR in FVA is significant for IRS, i.e. 10\%\ to more than 50\%\ and these results too may hold for other instruments.  

Technically we have shown how to deal with two and three factors in WWR.  This generalizes to any number of factors which can then be assessed, as here  with WWR2, for materiality. 

An alternative approach to WWR is via stress testing, i.e. calculate WWR against a list of market scenarios.  This is also model independent, but is typically  not included either in capital or nor pricing but only used for deal inclusion/exclusion, i.e. limits.  Stress testing complements pricing approaches such as the one described here.

Our analysis is on constant-composition portfolios as is standard.  This means that any hazard-driven exposure comes from credit portfolios where there is a  relationship between the counterparty hazard and reference hazards both in the portfolio and external to the portfolio.   Credit-driven exposure is subject to the same structural features as counterparty-driven-exposure so will be revealed in the term structure of exposure-default correlation.

 Our approach naturally calibrates to combined crisis and non-crisis data as demonstrated.  This can be used by any institution to estimate their multiplier for FRTB-CVA, or indeed for their regulator to do so, or ask that they do so.  This approach would naturally add to any QIS on FRTB-CVA revision --- and to reserves held by Product Control against WWR.  

Finally, our model independent approach can be used forensically to compare different WWR models on the same terms.  Models for comparison can simply be simulated and then use as calibration for this model independent approach.  In this way any WWR model that can produce simulated  paths of exposure, default and funding can be directly compared in terms of parameters directly relevant for CVA and FVA.  Just as different WWR models can be compared, so can different calibration assumptions for this model independent approach itself and other WWR models.

\section*{Acknowledgements}

The authors would gratefully like to acknowledge feedback from participants at the FIS Round-table (September 2019, Canary Wharf) and the 3rd Annual Derivatives Funding and Valuation conference (September 2019, Singapore).  Discussions with Lee Mcginty, Andrea de Vitis, Yousef El Otmani, Hayato Iida, Tom Cannon, and Nitin Adlakha were also useful.  Comments from the reviewers materially improved the content and precision of the paper.

\section*{Appendix 1: WWR Models}
\label{s:appendix}

The only liquid options on CDS are on credit indices, and these have a maximum maturity of less than one year (see Bloomberg screens).  Less than one-year volatilities are insufficient for XVA WWR except for worst-case models that do not require any credit volatilities at all \cite{haase2010model,glasserman2016bounding}.  

Partial list of WWR models:
\cite{finger2000toward,brigo2009counterparty,brigo2010bilateral,cespedes2010effective,haase2010model,uchida2010counterparty,buckley2011capturing,boukhobza2012cva,hull2012cva,pykhtin2012general,rosen2012cva,cherubini2013credit,li2013note,skoglund2013credit,bocker2014path,breton2014efficient,brigo2014nonlinear,ghamami2014stochastic,ballotta2015integrated,lee2015wrong,pang2015cva,ruiz2015optimal,xiao2015accurate,yang2015bilateral,baviera2016cva,glasserman2016bounding,brigo2017cva,feng2017cva,mbaye2017subordinated,memartoluie2017wrong,slime2017modeling,vrins2017wrong,kenyon2020model,chung2019cva,sakuma2020homotopy}.

\section*{Appendix 2: Equivalence of correlation, and correlation of expectations of samples}

Suppose we have two correlated random variables $x,y$ with variances $\sigma_x,\sigma_y$ and correlation $\rho$.

Now suppose we take sets of $n>0$ samples $\{x_i,y_i,\ i=1,\ldots n\}$ and calculate the correlations between the averages $\bar{x} = \sum x_i/n ,\ \bar{y} = \sum y_i/n$.
\begin{align*}
2\ \cov\lB\lB\sum x_i \rB \Big/n,\ \lB\sum y_i \rB \Big/n \rB 
=& \var\lB\lB\sum x_i \rB \Big/n + \lB\sum y_i \rB \Big/n \rB\\
& - \var\lB\lB\sum x_i \rB  \Big/n  \rB - \var\lB\lB\sum y_i \rB \Big/n \rB\\
=& \var \lB \sum x_i + \sum y_i  \rB  \Big/n^2 \\
&- \var\lB \sum x_i \rB \Big/n^2 + \var\lB\sum x_i \rB   \Big/n^2\\
=& \lB  2\rho \sigma_x \sigma_y/n  + \sigma_x^2/n + \sigma_y^2 /n\rB 
-  \sigma_x^2 /n - \sigma_y^2 /n\\  
=& 2 \rho \sigma_x \sigma_y /n
\end{align*}
Note that $x_i$ and $y_j,\ i\ne j$ have correlation zero, because they are from different samples.  Above we used $\var(\sum x_i/n) = n \sigma_x^2/n^2 = \sigma_x^2/n$, so re-using this we have
\begin{align*}
\text{correlation}(\bar{x},\bar{y}) &=  \frac{\rho \sigma_x \sigma_y / n}{ \sigma_x \sigma_y/n} \\
&= \rho
\end{align*}
So taking expectations of samples, then calculating correlation of the expectations, does not change the original correlation $\rho$.

\bibliographystyle{chicago}
\bibliography{xva-wwr}

\end{document}